\documentclass[preprint,showpacs,preprintnumbers,amsmath,amssymb]{revtex4}

\usepackage{graphicx}
\usepackage{dcolumn}
\usepackage{bm}

\begin{document}

\title{Blowup/scattering alternative for a discrete family of static critical solutions with various number of unstable eigenmodes.}

\author{Evgeny E.~Donets}
 \author{Edik A.~Hayryan}%
\author{Oksana I.~Streltsova}%
\affiliation{%
Joint Institute for Nuclear Research, 141980 Dubna, Russia
}%

\date{\today}

\begin{abstract}
Decay of regular static spherically symmetric solutions in the $SU(2)$ Yang-Mills-dilaton (YMd) system of equations under the independent excitation of their unstable eigenmodes has been studied self-consistently in the nonlinear regime. The considered regular YMd solutions form a discrete family and can be parametrised by the number $N=1,2,3,4...$ of their unstable eigenmodes in linear approximation. We have obtained strong numerical evidences in favour of the following statements: i) all static YMd solutions are distinct local threshold configurations, separating blowup and scattering solutions; ii) the main unstable eigenmodes are only those responsible for the blowup/scattering alternative; iii) excitation of higher unstable eigenmodes always leads to finite-time blowup; iv) the decay of the lowest $N=1$ static YMd solution via excitation of its unique unstable mode is an exceptional case because the resulting waves propagate as a whole without energy dispersion revealing features peculiar to solitons. Applications of the obtained results to Type-I gravitational collapse of massless fields are briefly discussed.
\end{abstract}

\maketitle

\section{\label{sec:level1}Introduction.}

In the last decade a considerable progress has been achieved in understanding of singular "blowup" solutions of nonlinear evolution equations with smooth initial data possessing finite energy. Some issues under consideration were inspired by pioneering work by Choptuik \cite{Ch1} on a gravitational collapse of massless field. This task has been developed in a number of new ways including gravitational collapse of nonlinearly interacting massless fields, such as Yang-Mills field and nonlinear sigma-models \cite{chbiztch},\cite{Lechner}, \cite{Gundlach}. Other related issues were motivated by recent studies on blowup in nonlinear wave equations \cite{Bizon5} - \cite{Aichelburg}. 

The nature of the threshold of singularity formation is of particular interest. This problem was studied analytically and, mostly, numerically and it was realized that there are at least two kinds of critical solutions which play a role of local thresholds, separating dispersing and singular solutions. 

The thresholds of first kind are self-similar solutions possessing one unstable mode. Critical solutions of this type can  be both discretely self-similar or continuously self-similar. Continuously self-similar thresholds are well-known 
solutions and their explicit form was found in some cases even analytically as solutions of system of ordinary differential equations after substitution of the "self-similar" anzats. Discretely self-similar critical solutions are more subtle objects and they were discovered first in numerical studies of Type-II 
gravitational collapse of massless fields \cite{Ch1}, \cite{Lechner}, \cite{Gundlach}, \cite{BizonSSul}. Black holes of arbitrary small mass  (Type-II gravitational collapse) can be produced in this way  and the critical discretely/continuously self-similar solution arises as a threshold, separating formation of naked singularity (zero-mass black hole) from dispersing solutions.

Another kind of the threshold configurations are static solutions 
with at least one unstable eigenmode. The well-known example of such 
a threshold is lowest $N=1$ Bartnick-McKinnon (BK) solution in a coupled system 
of Einstein-Yang-Mills equations \cite{BK}. This threshold solution 
provides Type-I gravitational collapse scenario with a finite mass gap 
in a spectra of produced black holes; the smallest black hole, 
formed in this way has a mass equal to $N=1$ BK static solution \cite{chbiztch}.

In this paper we focus on static threshold configurations. We consider spherically symmetric system of coupled $SU(2)$ Yang-Mills-dilaton (YMd) equations in $3+1$ Minkowski space-time. This system possess infinite number of finite energy  static regular  solutions  which are parameterized by the number of zeros $N=1,2,3,4...$ of relevant YM function \cite{LM}. All of these static solutions are unstable and have $N$ even-parity unstable eigenmode in a small linear perturbations approach. We have shown in one of our previous 
papers \cite{PRD} that this YMd system is supercritical in PDE terminology \cite{Klainerman} and all solutions of the corresponding evolutional Cauchy problem with "large enough" initial data result in finite time blowup.   

It is widely believed that a static unstable solution with one
unstable eigenmode could be a threshold configurations in 
supercritical PDE equations. In other words, these static solutions should have stable manifold of codimension-one and the instability results from a simple negative eigenmode. This conjecture was proven in some cases analytically 
\cite{Krieger} and numerically \cite{BizChTab}. It was shown numerically in various models that blowup/scattering alternative in a final solutions asymptotics results from the sign of the excited single unstable eigenmode \cite{BizChTab}, \cite{Szpak}.

The considered YMd system represents more general case because it
posessess an infinite family of the static solutions with 
$N=1,2,3,4,\ldots \infty$ unstable modes correspondingly.
A stable manifold of these solutions with  $N>1$ has a 
codimension more than one and few natural questions arise as a result: 
what is a role of higher static solutions with $N>1$; do they 
also relate to the distinct thresholds, separating blowup and 
scattering; and if so, what is the role of higher unstable 
modes in this blowup/scattering alternative?

To answer these questions, we systematically studied the decay of 
various lowest static YMd solutions under the independent excitations 
of all of their proper unstable eigenmodes. The results of this numerical 
studies are presented in Section 4. Section 2 contains basic 
notations and main field equations. In Section 3, we discuss the 
basic properties of the static intermediate attractors. Conclusions, 
discussion, and further open questions are summarized in Section 5.

\section{\label{sec:level2}Preliminary basic facts and equations.}

\hspace*{\parindent} A coupled system of Yang-Mills-dilaton (YMd) fields in $3+1$
Minkowski space-time is given by the action:
\begin{equation} \label{Vaction}
    S=\frac{1}{4\pi}\int\left(\frac{1}{2}(\partial\Phi)^2 -
     \frac{\exp[k\Phi]}{4g^2} F^{a\mu\nu} F^{a}_{\mu\nu}\right)\,d^3x\,dt.
\end{equation}

\noindent Here $\Phi$ is the dilaton field, $F^{a\mu\nu}$ is the Yang-Mills field, and $k$ and $g$ are dilaton and gauge coupling constants, respectively.

In the spherically symmetric case the dilaton field and the $SU(2)$ Yang-Mills potential
can be described in terms of two independent functions \begin{equation}
    \Phi=\Phi(t,r),\quad A^a_0=0\,,
\quad A^a_i= \epsilon_{aik}\frac{x^k}{r^2}\,
     \Big(f(t,r)-1 \Big).
     \label{v2}
\end{equation}
After substitution \eqref{v2} into \eqref{Vaction} and rescaling $\Phi\to\Phi/k$,
$ r\to (k/g)r$, $t\to (k/g)t$, $S\to g k S$ the dependence on two parameters  $k$ and $g$
effectively vanishes, and the integration over the angular variables gives finally the reduced action:
\begin{equation}
\label{action0I}
S=-\int\left[\frac{1}{2}r^2{\Phi_r}^2-\frac{1}{2}r^2{\Phi_t}^2+
e^{\Phi}\left({f_r}^2-{f_t}^2+\frac{(f^2-1)^2}{2r^2}\right)\right]\,dr\,dt.
\end{equation}
\noindent  The corresponding system of field equations has the form of a coupled nonlinear wave equations in the following way:
\begin{subequations}\label{evol0}
\begin{eqnarray}
&& f_{tt}+f_t\Phi_t-f_{rr}-f_r\Phi_r=\frac{f(1-f^2)}{r^2},  \label{4a} \\
&&\Phi_{tt}-\Phi_{rr}-\frac{2\Phi_r}{r}=-\frac{e^\Phi}{r^2}
\left(f_r^2- f_t^2 +\frac{(f^2-1)^2}{2r^2}\right). \label{4b}
\end{eqnarray}
\end{subequations}

Invariance of the field equations \eqref{evol0} with respect to the replacement
$f\to - f $ allows to put Yang-Mills function $f(t,0)=1$ at the origin without loss of generality. We look for finite energy solutions of \eqref{evol0} which are regular both at the origin and at the spatial infinity. These requirements provide boundary conditions
for the functions $\Phi(t,r)$ and $f(t,r)$ compatible with the field equations as follows:
\begin{eqnarray}
 f(t,r=0)=1,\quad  && f_r(t,r=0)=0,\quad \Phi(t=0,r=0)=\Phi_0, 
\quad  \Phi_r(t,r=0)=0; \label{boundary1} \\ 
\label{boundary2} \nonumber &&
  \lim_{r\rightarrow \infty}f(t,r)=\pm 1,\quad
  \lim_{r\rightarrow \infty}f_r(t,r)=0,\\[0.1cm]
&&\lim_{r\rightarrow \infty}\Phi(t,r)=\Phi_{\infty}, \quad
\lim_{r\rightarrow \infty }\Phi_r(t,r)=0.
\end{eqnarray}
Here $\Phi_0$ is an arbitrary finite constant and $\Phi_{\infty}$ is a constant, depending on $\Phi_0$ and on the total mass-energy of the
configuration. Note that according to the regularity requirement
Yang-Mills function should take its vacuum values $f=\pm 1$ at the origin and
at the spatial infinity.

The total mass-energy $E$ of the configuration described by the field equations
\eqref{evol0} is given by the energy functional:
 \begin{eqnarray}
 \label{Er}
M=E=\int\limits_0^{+\infty}\varepsilon(t,r) d r =
\int\limits_0^{+\infty}\left[\frac{1}{2}r^2{\Phi_r}^2 +
\frac{1}{2} r^2 {\Phi_t}^2 + e^{\Phi}\left({f_r}^2+{f_t}^2+\frac{(f^2-1)^2}{2r^2}\right)\right]d r,
\end{eqnarray}
where the energy density function $\varepsilon(t,r) = T_t^t(t,r)$ is introduced
in a standard way as the $tt$ component of the energy-momentum tensor $T_{\mu}^{\nu}$. Energy $E$ is unique {\it{on-shell}} conserved value for the system \eqref{evol0}: $dE/dt=0$.

 For the further sake it is useful to introduce radial momentum density function
$\rho(t,r)$ which is a $tr$ component of the energy-momentum tensor; in terms of the relevant functions $\Phi(t,r)$ and $f(t,r)$ it has form:
\begin{eqnarray}
\label{rho0}
\rho(t,r) = T^r_t=r^2 \Phi_t(t,r) \Phi_r(t,r) + 2e^{\Phi}\,f_t(t,r) f_r(t,r).
\end{eqnarray}
The total radial momentum $P^r \equiv \int \limits_0^{+\infty}T_t^r(t,r)dr$ is not conserved for the given system, since the action functional \eqref{action0I} contains explicitly radial variable $r$. However, at any given values of $t$ and $r$ this function $\rho(t,r)$ describes the total radial energy flux through the surface of sphere of radius $r$ and the sign of function $\rho(t,r)$ indicates the direction of energy propagation: positive sign of $\rho(t,r)$ corresponds to the ingoing (towards to the origin) energy flux and negative - to the outgoing energy flux.

Static ($t$~-independent) regular solutions of system  \eqref{evol0} with
the boundary conditions \eqref{boundary1}, \eqref{boundary2} are well-known since early 90th \cite{LM}. Similar to the Bartnick-McKinnon (BK) solutions of the SU(2) Einstein-Yang-Mills equations system \cite{BK}, there is an infinite discrete set of the regular static solutions $\{f_N(r)$, $\Phi_N(r)\}$ labeled by the total number $N$ of zeros of YM function  $f(r)$, $N=1,2,3,4,...\infty $, which were found numerically. The Yang-Mills functions $f_N(r)$ oscillate $N-1$ times, being bounded inside the strip $f_N(r) \in [-1,1]$, whereas the dilaton functions $\Phi_N(r)$ are monotonically growing.

The scale transformation:
\begin{equation}
 f \to f,\quad \Phi \to \Phi+\lambda,\quad r\to r\exp[\lambda/2],
\quad t\to t\exp[\lambda/2], \quad  \lambda=\mbox{const}  \label{scaletransf}
\end{equation}
does not change the field equations either. That is why without loss of generality one can put  $\Phi_0=\Phi(t=0,r=0)=0$ in \eqref{boundary1}.
The mass-energy takes a scale-dependent value and it transformed under 
the scale transformations \eqref{scaletransf} as:
\begin{equation}
 M \to \exp[\lambda/2]M.  \label{mscale}
\end{equation}

If the scale is chosen according to a $lim_{r \to \infty} \Phi_N(r) \to 0$, then the mass-energy of the corresponding static YMd solutions increases
with $N$ as follows: $M_1=1.60751$, $M_2=1.93112$, $M_3=1.98863$, $M_4=1.99814$,
$M_5=1.99970,$ $M_6=1.99995, ..., M_{\infty}=2.0$, saturating their upper limit $M_{\infty}=2.0$ as $N \to \infty$ \cite{LM}.

In terms of the linear perturbation analysis every such solution with $N$ zeros of YM function $\{f_N(r), \Phi_N(r)\}$ has $N$ even-parity unstable
 modes \cite{LM}.
Indeed, assume spherically symmetric perturbations of the regular static solutions $\{f_N(r), \Phi_N(r)\}$ of the following form:
\begin{equation} \label{voz}
       f(t,r)=f_N(r)+\epsilon \, e^{-\Phi_N(r)/2}\, v(r)\, e^{i\omega t} ,\quad
     \Phi(t,r) = \Phi_N(r) + \epsilon\, \frac{\sqrt{2}}{r} \, u(r)\, e^{i\omega t},
\end{equation}
where $\epsilon$ is a small parameter. This gives a standard matrix (2$\times$2) Sturm-Liouville eigenvalue problem with respect to the appropriately normalized eigenfunctions $\{u(r), v(r)\}$. The negative eigenvalues $\omega^2=\lambda<0$ correspond to the unstable eigenmodes. All the unstable eigenvalues  $\lambda_N^j, j=1,\ldots, N$ that correspond to the background solutions
with $N=1,2,3,4$ are presented in the following {\bf Table 1} in accordance  to our paper \cite{hep}. The corresponding eigenfunctions were also found and plotted in \cite{hep}, \cite{mpi} and we use them to determine the initial data for the Cauchy problem in Section 4.

\begin{table}
\caption{
\label{G1-tab1} 
Unstable eigenvalues $\{\lambda_N^i\}_{i=1}^N$ for the background static YMd solutions with $N=1,2,3,4$. 
        }
\begin{ruledtabular}
           \begin{tabular}{l|c|c|c|c}
     $N$ & $\lambda_N^1$& $\lambda_N^2$& $\lambda_N^3$& $\lambda_N^4$\\ 
                \hline
      1       & $-9.0566\times 10^{-2}$&                         &  &   \\
      2       & $-7.5382\times 10^{-2}$ & $-2.0742\times 10^{-4}$&  &   \\
      3       & $-4.9346\times 10^{-2}$ & $-1.4957\times 10^{-4}$&  $-1.9622\times 10^{-7}$ &    \\
      4       & $-4.3455\times 10^{-2}$ & $-5.9905\times 10^{-5}$&  $-1.3278\times 10^{-7}$& $\sim -10^{-9}$   \\
            \end{tabular}
            \end{ruledtabular}
\end{table}

  Now we proceed to another story inspired by the last decade developments in understanding of singularity formation for several classes  of nonlinear wave equations. It was shown in our previous paper \cite{PRD} that the considered system of Yang-Mills-dilaton equations \eqref{evol0} admits a hidden scale-invariant form. This observation allows to make definite predictions about possible late-time solution asymptotics in a nonlinear regime under the assumption of a well-posed evolutional Cauchy problem.

Indeed, if we decompose the dilaton function $\Phi(t,r)$ in the following way:
 \begin{equation}
\label{prphi}
    \Phi(t,r) = \phi(t,r) + 2 \ln r,
\end{equation}
the system of equations \eqref{evol0} rewritten in terms of functions  $f(t,r)$ and $\phi(t,r)$ becomes invariant under the global scale transformations of time and radial coordinate $t \to t/\lambda$ , $\quad r \to r/\lambda$. The mass-energy functional \eqref{Er} expressed in terms of the scale-invariant functions becomes
\begin{eqnarray}
E = \int dr \left(\frac{1}{2}r^2
\phi\,'^2+2r\phi\,\,'+2+\frac{1}{2}r^2\dot \phi^2
+r^2e^{\phi}\left( f\,'^2+\dot
f^2+\frac{(f^2-1)^2}{2r^2}\right)\right),  \quad \,\label{Etot1}
\end{eqnarray}
providing the homogeneous scale law for the mass-energy as
\begin{equation} \label{scalelaw}
  E\left[f(\frac{t}{\lambda},\frac{r}{\lambda}), \phi(\frac{t}{\lambda},\frac{r}{\lambda})\right]=
 \lambda^{\alpha} E[f(t,r), \phi(t,r)],\quad \alpha=+1.
\end{equation}

According to general expectations \cite{Klainerman} the degree $\alpha$
of scale parameter $\lambda$, occurring in the energy homogeneous scale law, defines the criticality class of the considered PDE equation, or system of equations. In particular, if $\alpha > 0$ the system is called {\it supercritical} as this takes place in our case and one should expect
singularity formation in the corresponding well-posed Cauchy problem in
a finite time and for all initial data, exceeding some threshold value.
One should also expect that a part of the solution shrinking to the origin evolves in certain universal way prior to the singularity formation. It
is believed that the mentioned universality is inspired by the existence of 
an attractor which these shrinking solutions attain for a while.
This attractors, as a rule, are described by self-similar solutions.
Exploring scale properties of the equations allows to search for an
explicit form of possible self-similar attractors.

In our case the revealed invariance of the solutions under the scale dilations
$$
f(t,r) \to f\left( \frac{t}{\lambda},
\frac{r}{\lambda}\right), \quad \phi(t,r) \to \phi\left(\frac{t}{\lambda},
\frac{r}{\lambda}\right)
$$
\noindent
allows to search for the solutions which effectively depend only on a particular combination of the independent variables - $t/r$, i.e. for
self-similar solutions as follows:
\begin{eqnarray}
 \label{avt-pred}
 f(t,r)= f(x), \quad   \phi(t,r)= \phi(x), \quad x = \frac{T-t}{r}.
\end{eqnarray}
The function $\phi(t,r)$ defined in \eqref{prphi} and the positive constant $T$  is transformed similarly to $r,t$ under the dilations. The constant $T$ could have the meaning of a blowup time - absolute value of the time in the evolution Cauchy problem when the singularity at the origin start to develop and the solution becomes nondifferentiable.

In terms of the functions $f(x)$ and $\phi(x)$ the system of PDEs \eqref{evol0} is transformed into a system of ordinary differential equations (ODEs).
This system of ODEs with the appropriate boundary conditions was considered in \cite{PRD} and the infinite discrete family of the desired
self-similar solutions was found. These self-similar solutions can be parametrized by $N=0,1,2,3,4,...+\infty$ - the number of zeros of the
Yang-Mills function $f(x)$. According to the performed linear perturbation analysis only the lowest self-similar solution with $N=0$ is stable.

And just this the lowest $N=0$ self-similar solution can pretend to be a global stable attractor in the evolutional Cauchy problem as was shown in  \cite{PRD}. The Cauchy problem for the system \eqref{evol0}, \eqref{boundary1}, \eqref{boundary2} has been studied numerically
for a wide range of smooth finite energy initial data. It turns out that if the initial data exceeds some threshold, the resulting solutions in a
compact region that shrinks to the origin attain the $N=0$ self-similar 
solutions. These solutions, being self-similar, further evolve in the universal
way in a finite time until the second derivative of the YM function at the origin starts growing infinitely what indicates on a singularity formation.

It should be noted that such a singularity formation (blowup) at the origin is not accompanied by energy concentration according to the energy
scale law \eqref{scalelaw}. This scale law describes the energy of a self-similar part of the solution as it evolves, shrinking to the origin, whereas
the total energy of the whole configuration \eqref{Er} is conserved. 
Indeed, the energy of self-similar part of the solution at
time $t \leq T$ ($T$~-blowup time) inside the past light cone of the point $t=T, r=0$ is equal to $M=E \sim 4(T-t)$ and vanishes as $t \rightarrow T$ \cite{PRD}.

The system dynamics determined by the near-threshold initial data in the mentioned Cauchy problem is a task of particular interest. In fact,
solutions with various initial data approaching their corresponding threshold values could attain for a while (prior to attain $N=0$ self-similar
attractor) some universal threshold configuration. In our case the $N=1$ static YMd solution is one of such threshold configurations that plays a role
of an intermediate attractor. We discuss this in more details in the next section. 

\section{Notes about static intermediate attractors.}

This section, as well as the previous one, contains rather preparatory material to the next section. We would like to show that the $N=1$ static YMd solution of the system \eqref{evol0}, \eqref{boundary1}, \eqref{boundary2} is a local intermediate attractor and it is also a threshold configuration separating blowup and scattering solutions. Following the conventional methodology we assume an isolated compact distribution of matter (YMd) fields as an initial data in the Cauchy problem \eqref{evol0}, \eqref{boundary1}, \eqref{boundary2}.
Further nonlinear evolution of the corresponding initially ingoing wave is a main subject of present section.

In this section we consider a compact distribution of YM field,
determined by the function $f(t=0,r)$ and by their partial
derivatives at the initial slice $t=0$. Then the initial profile
of the dilaton function $\Phi(t=0,r)$ is obtained by integration 
of the dilaton equation \eqref{4b} from the origin up to infinity
 assuming $\Phi(t, r=0)=0, \Phi(t=0,r)_{,t}=0, \Phi(t=0,r)_{,t,t}=0$ 
(see \cite{PRD} for more details ). The resulting value of the dilaton 
function at spatial infinity 
$\Phi(t,r\rightarrow \infty)= \Phi^{profile}_{\infty}$ is
a key parameter which defines the position of the virtual static
YMd solutions in this scale. Indeed, in order to study possible
attaining static solutions one should plot all their virtual
images at the same scale, i.e. with $\Phi_N(r \rightarrow \infty)=
\Phi^{profile}_{\infty}$ by using an appropriate shift of the dilaton
function and the scale transformations \eqref{scaletransf}. This was
done for all simulations discussed below.

Without loss of generality we assume two types of the initial YM profile described by the one-parametric kink-type and Gauss-type distributions as
follows:
 \begin{equation}
      f(t=0,r)= \frac{1-a\,r^2}{1+a\,r^2}, \quad \quad
      f(t=0,r) = 1-A\,r^2\exp\left[-\sigma(r-R)^2\right]\,, \label{p1}
\end{equation}
\noindent
where $\sigma$ is fixed and free parameters $a$, $A$ control the strength of the given initial data.

If the initial data are less than the threshold values in terms of $a_{cr}, A_{cr}$, than the initially ingoing waves get smoothly bounced near the
origin and are radiated away similarly to the solutions of linear wave equations. Positions of various YM and dilaton profiles at the moment of bouncing are plotted in the Figure 1 together with virtual images of the lowest $N=1,2$ static YMd solution at the appropriate scale. One can see that, as we are approaching the critical values of the parameters towards more strength initial data, the position of bounce is approaching from outside the virtual $N=1$ static YMd solution.

Dynamics of two various near-threshold solutions is presented on the Figure 2.  Behavior of the asymptotically scattered solution in
a vicinity of the virtual $N=1$ static solution is presented in the left column of the Figure 2. One can see that initially purely ingoing wave
attains the virtual static $N=1$ YMd solution for a while and after smooth bounce from it radiates away being transformed to the outgoing wave.
If the relevant parameter in the initial profile slightly exceeds the threshold value one has a blowup as a final asymptotics. The behavior of
the corresponding solution is presented in the right column of the Figure 2. The initially purely ingoing wave attains the virtual static $N=1$ YMd
solution for a while too, and then starts to collapse, attaining later on the universal $N=0$ self-similar solution which is a blowup precursor.

The time-dependent part of the energy functional for the considered solutions
\begin{eqnarray}
\label{ind}
\int\limits_0^{+\infty}\left[\frac{1}{2} r^2 {\Phi_t}^2 + e^{\Phi} f_t^2 \right]d r,
\end{eqnarray}
has a local minima at some instant of time for all solutions which attain the intermediate attractor. Gradual change of the initial parameters $a$
and $A$ in a small vicinity of their threshold values allows to minimize this integral over all range of the initial data parameters. The corresponding
positions of the solution profiles (plotted at the moment when the integral has a minimum) converge to the limiting (static) threshold configuration
separating asymptotically scattered and blowup solutions. It is not surprising that the  $N=1$ static YMd solution is occurred to be a such
limiting threshold configuration. This step-by-step shooting numerical strategy allows to approach the threshold value of the relevant parameter
in the initial data with precision about $10^{-11}$.

The mass-energy of the static intermediate attractor is an
absolute lower bound for the mass-energy of all solutions of the
ingoing wave type which could attain this attractor. This
(unreachable) limit corresponds to the vanishing integral
\eqref{ind}. Hence, in order to have a chance to attain a static
intermediate attractor, the mass-energy of the initial field
configuration must exceed the mass-energy of the corresponding
virtual static solution (calculated at the appropriated scale
determined by the $\Phi^{profile}_{\infty}$) which pretends to be
an attractor. In this case a considerable part of the energy
surplus should radiate away prior the solution attains the static
attractor. The observed formation of an outgoing energy flux could
serve as a useful numerical tool to resolve a possible attaindance to
the intermediate attractor. The bottom plots of the Figure 2
illustrate this statement: the negatively defined parts of the
radial momentum density function $\rho(t,r)$ testify to occurrence of
the outgoing energy fluxes which are formed prior the solutions
attain the static $N=1$ intermediate attractor.

To summarize: at least for those solutions which temporary attain the virtual $N=1$ static YMd solution the last one plays a role of a local threshold configuration separating scattering and blowup solutions. This is an expected
behavior for the supercritical systems admitting static solutions with one unstable eigenmode. Note that this is the most clear case,
since due to an evident transversality, a stable manifold of the $N=1$ static solution has codimension $1$. Under the circumstances it was possible to observe convergence to the $N=1$ static threshold configuration starting with the arbitrary chosen simple kink-type or Gauss-type initial data at spatial infinity.

Situation becomes more complicated if we consider higher static YMd
solutions with $N>1$ which have more than one unstable eigenmodes. 
Stable manifold of these solutions with the $N>1$ has codimension~$>1$ 
as well and in this case, for a number of reasons, it is difficult to create initial data at spatial infinity which could control (by the fine tuning of the governing parameters) the possible attaindance to the higher $N>1$ static solutions pretending to be distinct threshold configurations.

Instead, it is reasonable to start directly with this unstable static solutions and study their decay via unstable eigenmodes excitation.
The role of various negative eigenmodes studied from the viewpoint of blowup/scattering alternative is a main task of this paper and is considered
in the next section.
 
 \section{Decay of static Yang-Mills-dilaton solutions and their asymptotics.}
 
 We consider the coupled system of the YMd nonlinear wave equations \eqref{evol0} with the boundary conditions \eqref{boundary1}, \eqref{boundary2}. The
 initial data at $t=0$ are determined by the static solution $\{f_N(r), \Phi_N(r)\}$
 ($N$ is fixed) perturbed by their proper unstable eigenfunctions $\{v_N^i(r), u_N^i(r)\}$, ($i$ is also fixed, $1 \leq i \leq N$) with the corresponding
 eigenvalue $\lambda_N^i=\omega^2<0$ according to \eqref{voz} and {\bf{Table 1}} as follows:
 \begin{equation} \label{incond1}
        f(t=0,r)=f_N(r)+\epsilon \, e^{-\Phi_N(r)/2}\, v_N^i(r),\quad
      \Phi(t=0,r) = \Phi_N(r) + \epsilon\, \frac{\sqrt{2}}{r}\, u_N^i(r)\ ,
 \end{equation}
 where $\epsilon$ is small parameter. Time derivatives at the initial time slice $t=0$ are obtained from the \eqref{voz} as follows:
  \begin{equation} \label{incond2}
        f_t(t=0,r)=\epsilon \,\sqrt{-\lambda_N^i}  e^{-\Phi_N(r)/2}\, v_N^i(r),\quad
      \Phi_t(t=0,r) = \epsilon\,\sqrt{-\lambda_N^i} \frac{\sqrt{2}}{r}\, u_N^i(r)\ .
 \end{equation}
 The normalized eigenfunctions $\{v_N^i(r), u_N^i(r)\}$ and the eigenvalues $\lambda_N^i$ ({\bf{Table 1}}) are taken from our previous papers \cite{hep}, \cite{mpi}. Some lowest eigenfunctions $\{v_N^i(r), u_N^i(r)\}$ are plotted in Fig.3 (top), as they enter into the relevant combinations in the formula \eqref{incond1}:
 \begin{equation} \label{vf}
        V^i_f(r)= e^{-\Phi_N(r)/2}\, v_N^i(r),\quad
      V^i_{\Phi}(r) =\frac{\sqrt{2}}{r}\, u_N^i(r).
 \end{equation}
 The background functions $f_N(r)$, $\Phi_N(r)$ and their unstable
 eigenfunctions $v_N^i(r), u_N^i(r)$ have the following behavior:
 \begin{eqnarray}
 \label{uvbound} 
 && f_N(r=0)=1,\quad f_N '(r=0)=0,\quad f_N(r \to
 \infty)=\pm 1,\quad f_{N} '(r \to \infty)=0, \\
 && \Phi_N(r=0)=0,\quad
 \quad \Phi_{N} '(r=0)=0,\quad \Phi_N(r \to \infty)=\tilde{\Phi}_N,\quad
  \Phi_{N} '(r \to \infty)=0 , \\
 && v_{N}^{i} (r=0)=0,
 \quad {v_{N}^{i}}'(r=0)=0,
 \quad v_{N}^{i} (r \to \infty)=0,
 \quad {v_{N}^{i}}'(r \to \infty)=0,\\
 && u_N^i(r=0)=\tilde{u}_N^i, \quad {u_N^i}'(r=0)=0,\quad u_N^i(r \to
 \infty)=0,\quad {u_N^i}'(r \to \infty)=0 ,
 \end{eqnarray}
 where $\tilde{\Phi}_N$ and $\tilde{u}_N^i$ are the obtained constants.
 
 As a result, the boundary conditions for the considered evolutional Cauchy problem are summarized as follows:
 \begin{eqnarray}
 \label{finalboundary1}
  f(t,r=0)=1,\quad  && f_r(t,r=0)=0,\quad \Phi(t=0,r=0)=\tilde{u}_N^i,
 \quad  \Phi_r(t,r=0)=0; \\ 
 \label{finalboundary2} \nonumber &&
   \lim_{r\rightarrow \infty}f(t,r)=\pm 1,\quad
   \lim_{r\rightarrow \infty}f_r(t,r)=0,\\[0.1cm]
 &&\lim_{r\rightarrow \infty}\Phi(t,r)=\Phi_N, \quad
 \lim_{r\rightarrow \infty }\Phi_r(t,r)=0.
 \end{eqnarray}
 
 Note that the value of the dilaton function at the origin is defined by the boundary conditions at the initial time only: $\Phi(t=0,r=0)=\tilde{u}_N^i$.
 However, on each other slice $t>0$ this value is a free parameter that evolves according to the equations \eqref{evol0}. In other words,
 we put free boundary condition for the dilaton function at the origin $\Phi(t>0,r=0)$. At the same time we keep the value of the dilaton fixed at infinity equal to its initial value $\lim_{r\rightarrow \infty}\Phi(t \ge 0,r)=\Phi_N$ that provides the same scale during the evolution.
 
 Using the normalization condition for the unstable eigenfunctions $\{v_N^i(r), u_N^i(r)\}$ \cite{hep} one can obtain a useful formula for the mass-energy shift
 caused by adding eigenfunctions $\{v_N^i(r), u_N^i(r)\}$ to the background static solutions in accordance with \eqref{incond1}
 \begin{equation} \label{energP}
 E= E_N+2\epsilon^2 \left|  \lambda_N^i \right| ,
 \end{equation}
 where $E_N$ is the mass-energy of the corresponding static solution $\{f_N(r), \Phi_N(r)\}$ and  $\epsilon$ is a free small parameter. As we will see below,
 the sign of the free small parameter $\epsilon$ plays a key role for the late-time solution asymptotics. However, it enters squared into the mass-energy
 shift formula \eqref{energP} in the framework of the considered perturbation scheme.
 
 So we have a well-posed Cauchy problem of mixed type (both,
 initial and boundary conditions are imposed), which has been
 solved numerically at various initial conditions with the help of an
 adaptive mesh refinement of the finite difference scheme that preserves
 the total energy during the evolution (see \cite{PRD} for further
 details). The semi-infinite interval $r \in [0,\infty)$ was
 replaced by an appropriate finite interval $r\in [0,R_\infty]$ for the
 numerical reasons.

  \subsection{Decay of the $N=1$ static solution.}
 
 Let us consider the $N=1$ static solution  perturbed by its single unstable eigenmode with $\lambda_1^1=-9.0566\times 10^{-2}$ as initial data:
 \begin{equation} \label{incond11}
        f(t=0,r)=f_1(r)+\epsilon \, e^{-\Phi_1(r)/2}\, v_1^1(r),\quad
      \Phi(t=0,r) = \Phi_1(r) + \epsilon\, \frac{\sqrt{2}}{r}\, u_1^1(r)\ ,
 \end{equation}
 \begin{equation} \label{incond21}
        f_t(t=0,r)=\epsilon \,\sqrt{-\lambda_1^1}  e^{-\Phi_1(r)/2}\, v_N^i(r),\quad
      \Phi_t(t=0,r) = \epsilon\,\sqrt{-\lambda_1^1} \frac{\sqrt{2}}{r}\, u_1^1(r)\ .
 \end{equation}
 The results of various simulations with different values of $\epsilon$ allow us to conclude that the late-time asymptotics of the solutions depends on the sign of the parameter $\epsilon$ only. In fact, if the parameter $\epsilon<0$ is negative, the solutions evolve for a while in the linear regime \eqref{voz}, then start to collapse and later on they attain the stable $N=0$ self-similar attractor. Finally, the singularity develops at the origin in a finite time $T$. The absolute value of $T$ depends on the value of $\epsilon$ in the initial data. All such solutions with various initial $\epsilon<0$ have the universal self-similar late-time asymptotics with the effective dependence on the difference of the time $t$ and the absolute blowup time $T$ as $r/(T-t)$. 
 
 To recognize the self-similar nature of the dilaton function, one should extract its self-similar part $\phi(t,r)$ according to the \eqref{prphi}, which corresponds to the frame appropriately shifted and rotated at the angle $\beta=arctg(-2)$ back clockwise (see \cite{PRD} for details).
 
 Typical profiles of the YM and the dilaton functions at various times are shown in Fig.~4. One should indicate the fact that the corresponding
 radial momentum density function \eqref{rho0} occurred to be always positively defined, starting with the initial time $t=0$ (see Fig.~3, bottom) until the blowup. This means that there is no any outgoing energy flux at any time during the evolution.
 
 If $\epsilon>0$ the solutions evolve in the linear regime \eqref{voz} for a while, then the nonlinear waves start moving towards a large $r$ and
finally one has a purely outgoing linear waves as a late-time solutions asymptotics. Indeed, as the outgoing wave propagates towards
 $r \rightarrow +\infty$, the influence of the nonlinear terms in the right hand sides of the equations \eqref{evol0} becomes negligible and one
 deals with the linear system of wave equations as follows:
 \begin{eqnarray}\label{linear}
 && f_{tt}+f_t\Phi_t-f_{rr}-f_r\Phi_r=0,   \\
 &&\Phi_{tt}-\Phi_{rr}-\frac{2\Phi_r}{r}=0.
 \end{eqnarray}
 The solutions of the linear system \eqref{linear} have the standard form
 \begin{eqnarray}
  \label{linsol}
  f(t,r)= f_{lin}(t-r), \quad   \phi(t,r)= \frac{\phi_{lin}(t-r)}{r},
 \end{eqnarray}
 where $f_{lin}(t-r)$ and $\phi_{lin}(t-r)$ are arbitrary functions.
 The asymptotical profiles of the relevant YM $f_{lin}(t-r)$ and the dilaton $\phi_{lin}(t-r)$ functions are shown in Fig.~5 in the limit $\epsilon \to 0$. One should stress that similarly to the previous case $\epsilon<0$, for $\epsilon>0$ the only purely outgoing energy flux was observed at any time of the evolution. In this respect the decay of the $N=1$ static YMd solutions via their unstable eigenmode excitation is rather exceptional. 
 
 \subsection{Decay of the $N=2$ static solution.}
 
 Now we consider first the static $N=2$ solution perturbed by its main unstable eigenmode with $\lambda_2^1=-7.5382 \times 10^{-2}$ and with the eigenfunctions $\{v_2^1(r), u_2^1(r)\}$ in the initial data \eqref{incond1}, \eqref{incond2}. 
 
 The obtained results occurred to be qualitatively the same as in the previously described case of the static $N=1$ solution decay.  For $\epsilon<0$ the corresponding solutions start to collapse and later on they attain the $N=0$ self-similar attractor that leads to finite-time blowup, see Fig.~6. However, in contrast to $N=1$ decay some outgoing energy flux is formed at the initial time as a sequence of the unstable eigenfunction shape at its outer side 
 (see Fig.~3, bottom). One can observe that this outer part of the solution develops soon and then starts to propagate towards the spatial infinity separately on the main part of the solution (Fig.~6: bottom, right).
     
  For $\epsilon>0$ the solutions radiate away to the spatial infinity. Fig.~7 shows the typical pictures of such solutions.
 
 If we consider excitation of the second unstable eigenmode with $\lambda_2^2=-2.0742 \times 10^{-4}$ and with the eigenfunctions $\{v_2^2(r), u_2^2(r)\}$ in the initial data \eqref{incond1}, \eqref{incond2}, the late-time solutions asymptotics occurred to be independent on the sign of the parameter $\epsilon$. Indeed, after some small oscillations near the static $N=2$ configuration (see Fig~. 8, 9) both solutions start to collapse, then they attain the $N=0$ self-similar attractor that leads to a finite-time blowup. In other words, the excitation of second unstable eigenmode in the $N=2$ static solution does not provide a blowup/scattering alternative for the late-time solutions asymptotics. As we will show below, the same is true for all higher unstable eigenmodes of other static solutions with $N>1$: the independent excitation of higher unstable eigenmodes always leads to the finite-time blowup.
 
 Note that small but detectable outgoing energy flux was observed in both cases $\epsilon>0$, $\epsilon<0$ if the second unstable mode was excited. Similar to the main unstable mode excitation the outgoing pulse is formed already at the initial time owing to the unstable eigenfunctions shape at their outer sides. 
 
  \subsection{Decay of the $N=3,4$ static solutions.}
 
 The static $N=3$ and $N=4$ solutions have 3 and 4 different unstable eigenmodes, respectively.
 
 An excitation of the main unstable modes with $\lambda_3^1=-4.9346
 \times 10^{-2}$ for the $N=3$ YMd static solution and with
 $\lambda_4^1=-4.3455 \times 10^{-2}$ for the $N=4$ static solution
 provide a blowup/scattering alternative where the sign of $\epsilon$ is a key parameter, similar to the  $N=1,2$ static YMd solutions decay. 
 
 Similarly to the static $N=2$ solution decay, in the case of negative values of $\epsilon$, which leads to blowup, a small but detectable outgoing energy flux was observed arising at the initial moment of time, see Fig.~10.
 
Along the same lines, the independent excitations of the higher unstable modes (with $\lambda_3^2=-1.4957 \times 10^{-4}$,
$\lambda_3^3=-1.9622 \times 10^{-7}$ for the $N=3$ static solution and with $\lambda_4^2=-5.9905 \times 10^{-5}$,
$\lambda_4^3=-1.3278 \times 10^{-7}$, $\lambda_4^4=-1. \times 10^{-9}$ for the $N=4$ static solution) lead to the finite-time blowup
independently on the sign of the parameter $\epsilon$ in the initial data. As a result, the asymptotical behavior of these solutions is independent
on the sign of $\epsilon$, quite similar to the $N=2$ static solutions decay, discussed previously. Similarly to the decay of the $N=2$ static solution via their second unstable mode excitation, a small but detectable outgoing energy flux was observed in both cases $\epsilon>0$, $\epsilon<0$ if the highest unstable modes have been excited.

 We have also studied the possibility of attaining some lower $N-1, .. 1$ YMd static solutions following the decay of the static YMd solutions with  $N$ zeros of the YM function ($N=2,3,4$) in both ways (blowup/scattering). One can conclude that these solutions never attain other possible
 intermediate attractors of the low mass-energy. 
 
 It seems reasonable that the decay of higher $N>4$ static YMd solutions is qualitatively similar to the studied ones with $N=1,2,3,4$. This allows us to draw the following conclusions.  

   \section{Conclusions and discussion.}
 
 We obtained strong numerical evidences in favour of the following statements:
 
 - All static YMd solutions are distinct local threshold configurations, separating blowup and scattering solutions.
 
 - The main unstable eigenmodes are only those responsible for the blowup/ scattering alternative in the final solution asymptotics for all static YMd solutions.
 
 - A sign of the small parameter $\epsilon$ in \eqref{incond1} is a key parameter, providing blowup/scattering alternative when the main unstable eigenmodes are excited.
 
 - Excitation of higher unstable eigenmodes always leads to finite-time blowup independently on the sign of $\epsilon$.
 
 - Decay of the $N=1$ static YMd solution via excitation of its unique unstable eigenmode is an exceptional case because the resulting waves revealing features peculiar to solitons. It was shown that such waves propagate in both radial directions as a whole without energy dispersion.

 Following the Bizon-Tabor conjecture \cite{Bizon6}, we believe that singularity formation key properties are not specific features of definite nonlinear evolution equation, but rather are universal characteristics of a criticality class which this equation belongs to. In this way the results obtained for blowup can be valid as well for all supercritical systems of the same type and can shed new light on the old problems inspired, in particular, by the last decade studies of black hole formation as a result of the collapse of massless matter fields.
  
 Let us remind that if gravity is included, say in the Einstein-Yang-Mills (EYM) system of spherically symmetric equations \cite{chbiztch}, a smallest black hole mass formed in Type-I gravitational collapse occurs to be exactly equal to the mass of the lowest unstable $N=1$ Bartnick-McKinnon (BK) solution.  It is possible to take for granted that $N=1$ BK solution is a local threshold configuration which separates dispersing and collapsing solutions. Thus, the smallest black hole in Type-I collapse is formed if the unique unstable eigenmode of $N=1$ BK solution is excited with an appropriate sign of  $\epsilon$ in a limiting case as $\epsilon \to 0$. If these statements are accepted, then a natural question arises of why all matter which forms $N=1$ BK solution appears inside black hole event horizon and no any outgoing energy flux is produced. In fact, tales of outgoing pulses formed due to backscattering of ingoing collapsing waves on a space-time curvature are most general and universal sources of information about collapse details for a distant observer.
In addition to this a various basic processes inside black holes,
such as a mass inflation \cite{Poisson}, \cite{Belinskii} phenomenon, are also inspired by the backscattering of ingoing matter pulses.  In the case under consideration an absence of such outgoing energy flux during $N=1$ BK solution decay is very unusual and interesting.   
 
  On the basis of the obtained conclusions, we believe that any supercritical system (with gravity included or not) has similar origin of this phenomenon.  As we have shown, decay of the lowest static solution in YMd system via its single unstable eigenmode excitation produces exceptional nonlinear soliton-like wave, which can propagate in both radial directions as a whole without backscattering of the energy. Accepting the universality conjecture it is possible to assume that similar soliton-like wave is formed in Einstein-Yang-Mills system that naturally explains the equality of the smallest black hole mass to the mass of the $N=1$ BK solution. We plan to study soliton-like behavior of such solutions in more detail in a separate paper. 
 
 On the other hand, the collapse decay of higher $N>1$ static unstable solutions is always accompanied by some energy flux to spatial infinity. Thus, the question about lower bound of mass-energy kink, attaining some distinct $N>1$ static threshold configuration prior to blowup, is more subtle. We plan to consider this problem separately as well.    
 
 \begin{acknowledgments}
 Discussions with V.P. Gerdt and A.V. Glagyshev are greatly acknowledged. This research was supported in part by RFBR grant 05-01-00645, RFBR grant 06-01-00530 and  RFBR-BRFBR  grants 06-01-81014, 04-01-81011.
 \end{acknowledgments}




\newpage
\begin{figure}[!ht] 
 \label{fig:01}
    \includegraphics[width=16cm]{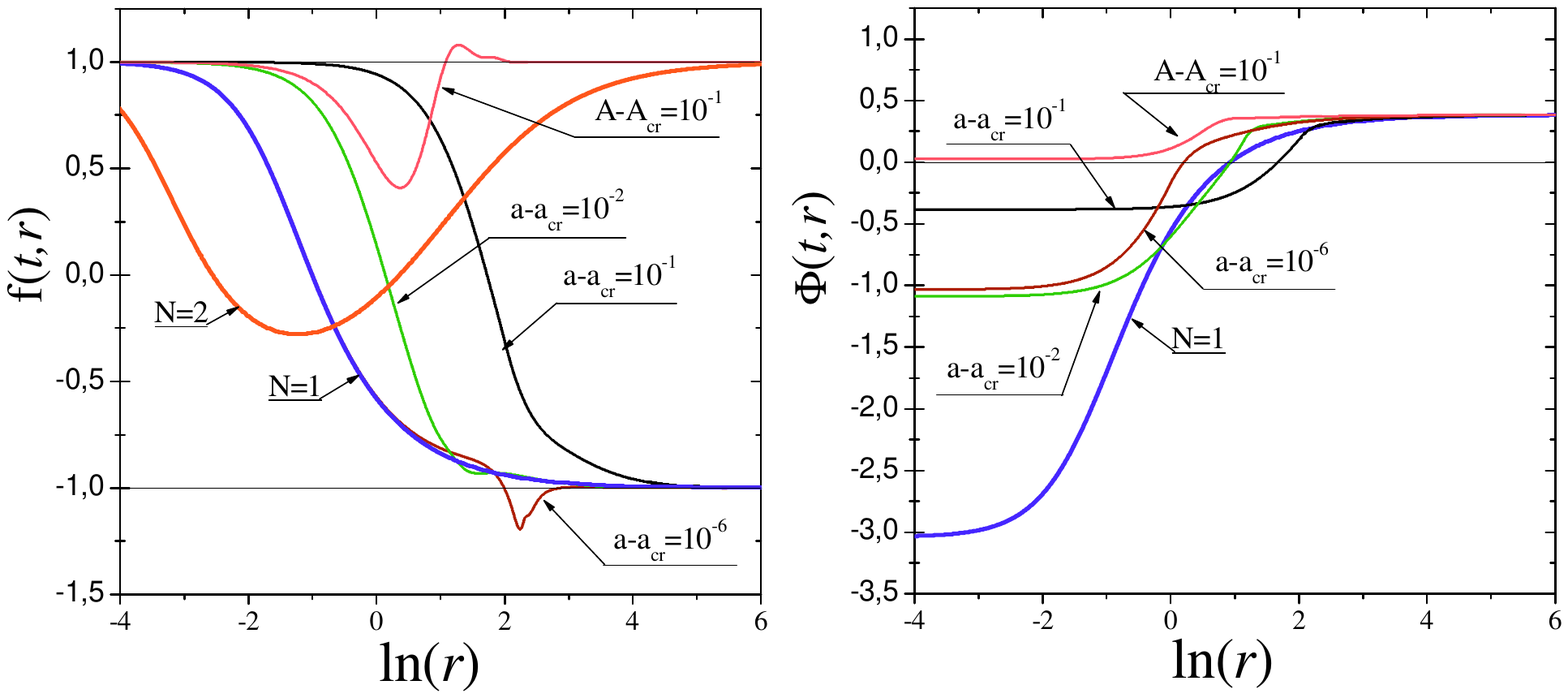} 
     \vspace*{-14cm} 
      \caption{Positions of YM functions $f(t,r)$ (left) and the dilaton 
functions $\Phi(t,r)$ (right) with various initial data, shown in the moment of bouncing. Virtual positions of the $N=1,2$ static YMd solutions are also plotted in the same scale. One can see that the virtual $N=1$ static YMd solution plays a role of border line in space; all dispersing solutions can not penetrate this border line being bouncing outside. 
}
\end{figure}


\newpage

\begin{figure}[!ht] 
 \label{fig:6}
    \includegraphics[width=17cm]{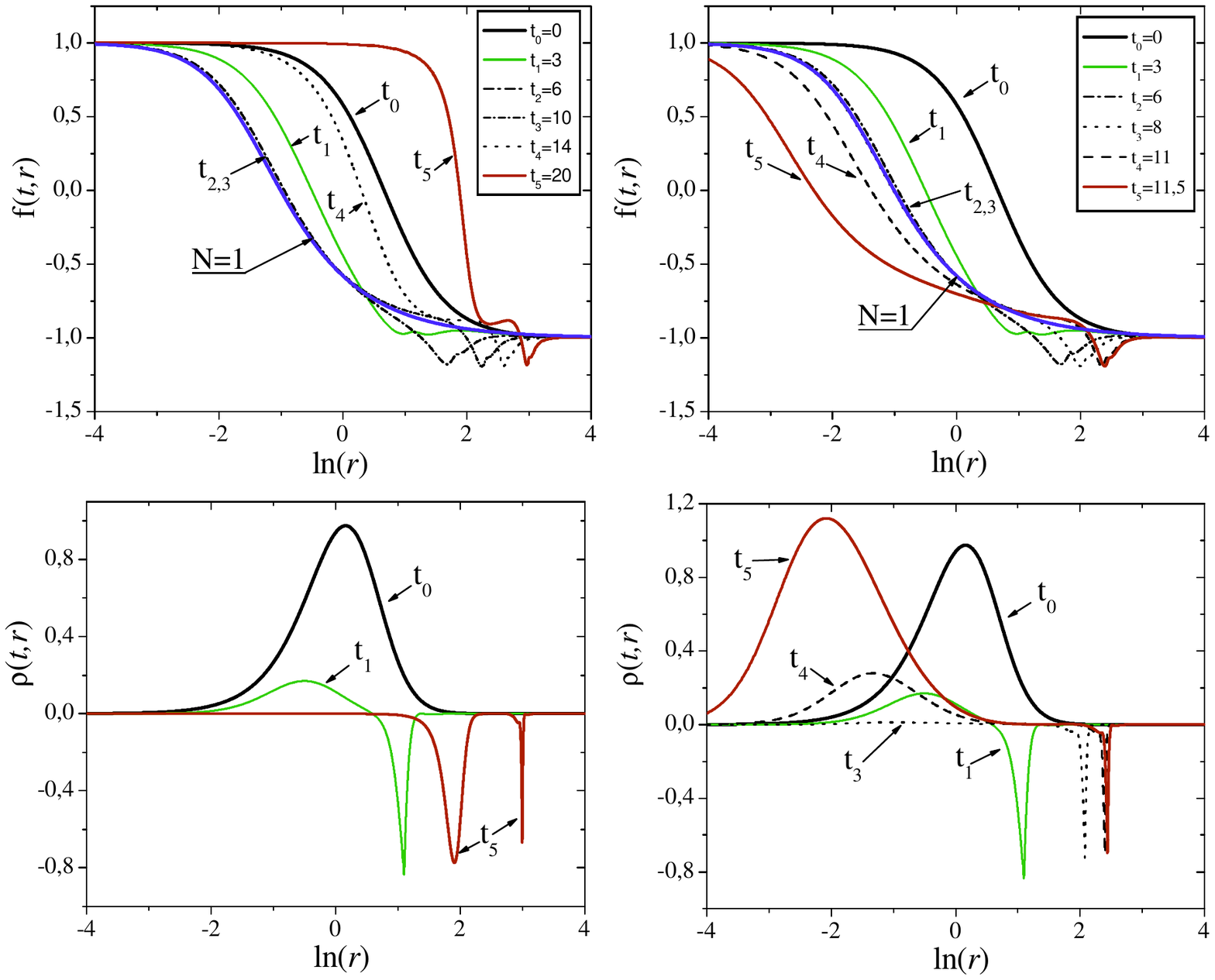} 
         \vspace*{-10cm} 
      \caption{
Evolution of the kink-type initial data for the YM function $f(t,r)$ with the key parameter values $a$, closed to the threshold one: $a<a_{crit.}$ (left,top), $a>a_{crit.}$ (right, top). The mass-energy access which is approximately equal to the difference between the kink mass-energy and the mass-energy of the $N=1$ static solution, is radiated away towards $r \rightarrow \infty$ before the solution attains this intermediate attractor. The plot of the corresponding radial momentum density function $\rho(t,r)$ is also presented (bottom): the negative $\rho(t,r)$ corresponds to the outgoing energy flux.  
}
\end{figure}

\begin{figure}[!ht] 
 \label{fig:7}
    \includegraphics[width=17cm]{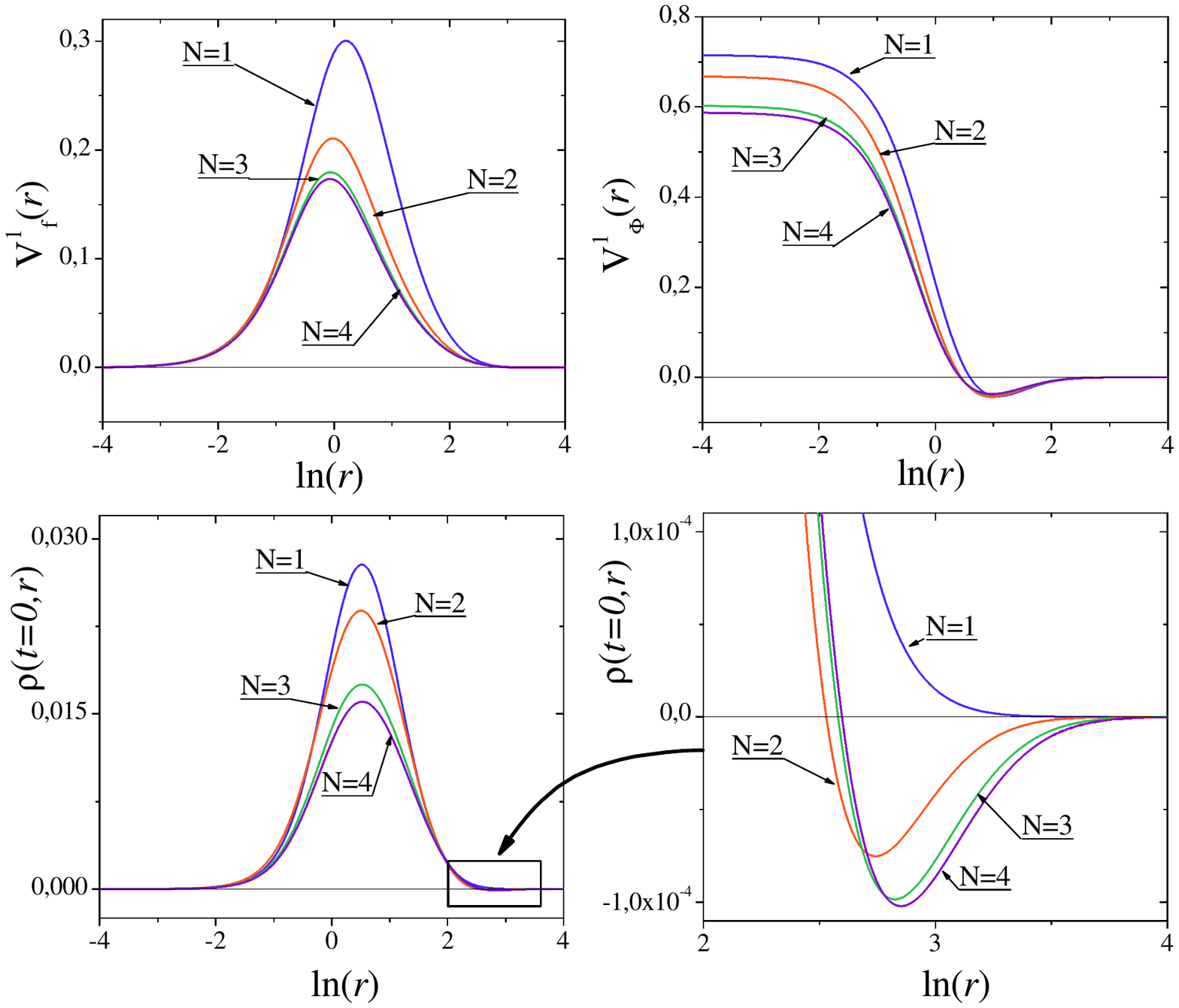} 
          \vspace*{-10cm} 
      \caption{
The  main unstable eigenfunctions entering in the relevant combination \eqref{vf} for the static YMd solutions with $N=1,2,3,4$ (top). The radial momentum density function $\rho(t=0,r)$ at the initial moment $t=0$ in the Cauchy problem \eqref{incond1}, \eqref{incond2} for the static YMd solutions with $N=1,2,3,4$ perturbed by their main unstable eigenfunctions (bottom , left) with $\epsilon<0$. The same is plotted (bottom, right) with the increased resolution in a vicinity of the right edge of the pulse. One can see that the only $N=1$ perturbed static YMd solution has a definite sign of radial density function $\rho(t=0,r)$ everywhere.  
}
\end{figure}

\newpage
\begin{figure}[ht!] 
 \label{fig:1}
   \centering 
\includegraphics[width=16cm]{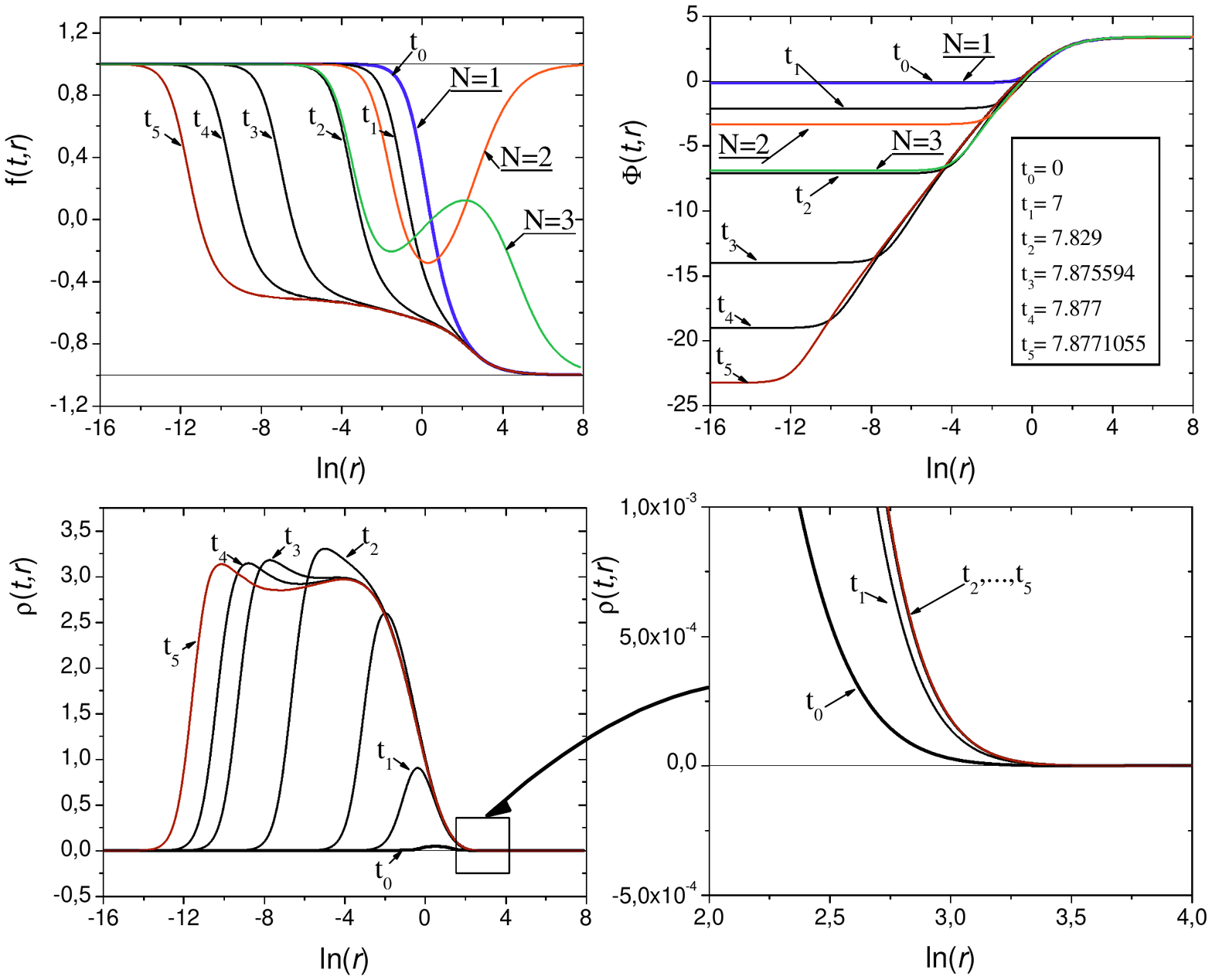} 
      \vspace*{-10cm} 
 \caption{The decay of the $N=1$ static solution along its single unstable eigenmode with $\lambda_1^1=-9.0566\times 10^{-2}$ and $\epsilon=-0.2$ which leads to blowup. The YM $f(t,r)$ profiles (top, left) and the dilaton profiles $\Phi(t,r)$ (top, right) at the various moments of time are plotted versus the logarithm of $r$. Virtual positions of the $N=2,3$ static YMd solutions at the same scale are also shown. At the later time, prior to blowup, the solution attains the stable self-similar $N=0$ profile. The radial momentum density function $\rho(t,r)$ (bottom, left) and the energy density function $\epsilon(t,r)$ (bottom, right) are displayed at the same moments of time, 
which exhibits the purely ingoing energy flux.}
\end{figure}

\newpage

\begin{figure}[h!] 
\label{fig:2}
\includegraphics[width=16cm]{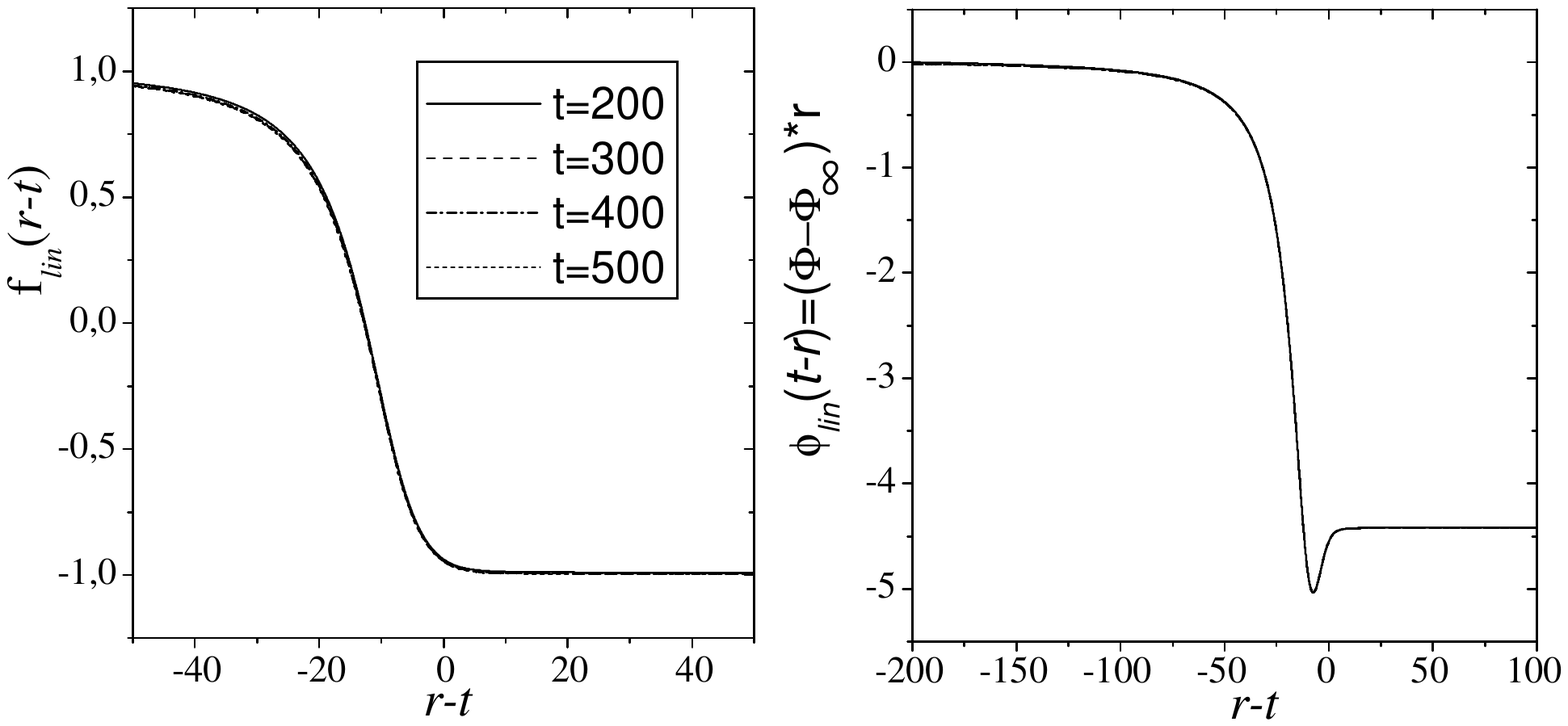} 
      \vspace*{-13cm} 
    \caption{
The asymptotical profiles of the YM function $f_{lin}(t,r)$ (left) and profiles of the normalized dilaton function $\phi_{lin}(t-r)=(\Phi(t,r)-\Phi_{\infty})\times r$ ($\Phi_{\infty}=3.35879184$ is the asymptotical value of the dilaton function for the $N=1$ static solution) (right) are displayed at later moments of time versus retarded null coordinate $r-t$. These asymptotical profiles are the result of the $N=1$ static solution decay along its single unstable eigenmode with $\lambda_1^1=-9.0566\times 10^{-2}$ in the limit $\epsilon \to 0$.
}
\end{figure}

\newpage

\begin{figure}[h]
 \label{fig:3}
    \includegraphics[width=16cm]{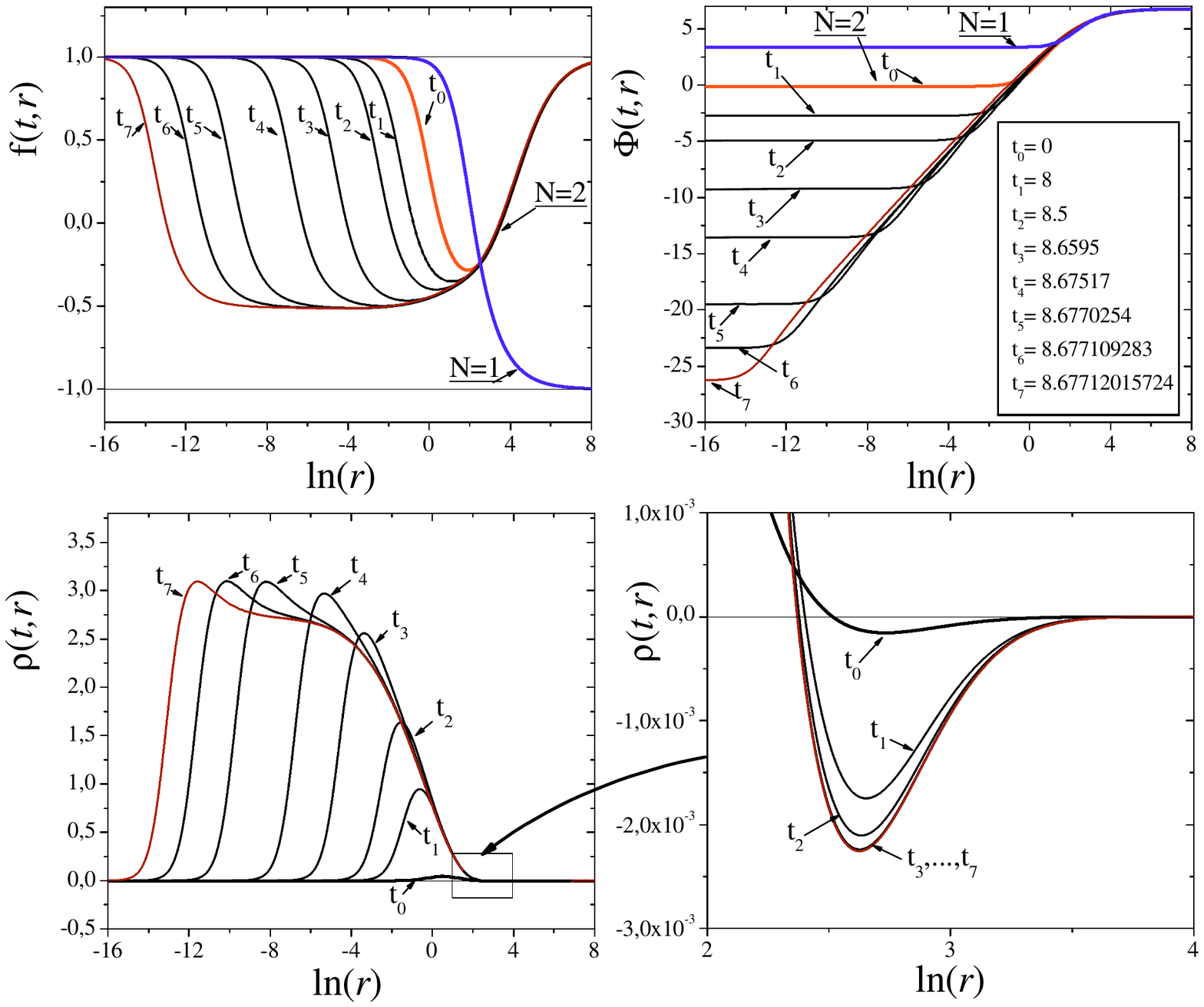} 
          \vspace*{-9cm} 
    \caption{
The decay of the $N=2$ static solution along its main unstable eigenmode 
with $\lambda_2^1=-7.5382 \times 10^{-2}$ and $\epsilon=-0,2$, which leads to blowup. The YM $f(t,r)$ profiles (top, left) and the dilaton profiles $\Phi(t,r)$ (top, right) at the various moments of time are plotted versus the logarithm of $r$. At the later times, prior to blowup, the solution attains the stable self-similar $N=0$ profile. The radial momentum density function $\rho(t,r)$ (bottom: left, right) is displayed at the same moments of time. 
The negative values of the function $\rho(t,r)$ (bottom, right) indicate an 
outgoing energy flux, which can be detected at spatial infinity.
}
\end{figure}


\newpage

\begin{figure}[h!] 
\label{fig:4}
    \includegraphics[width=17cm]{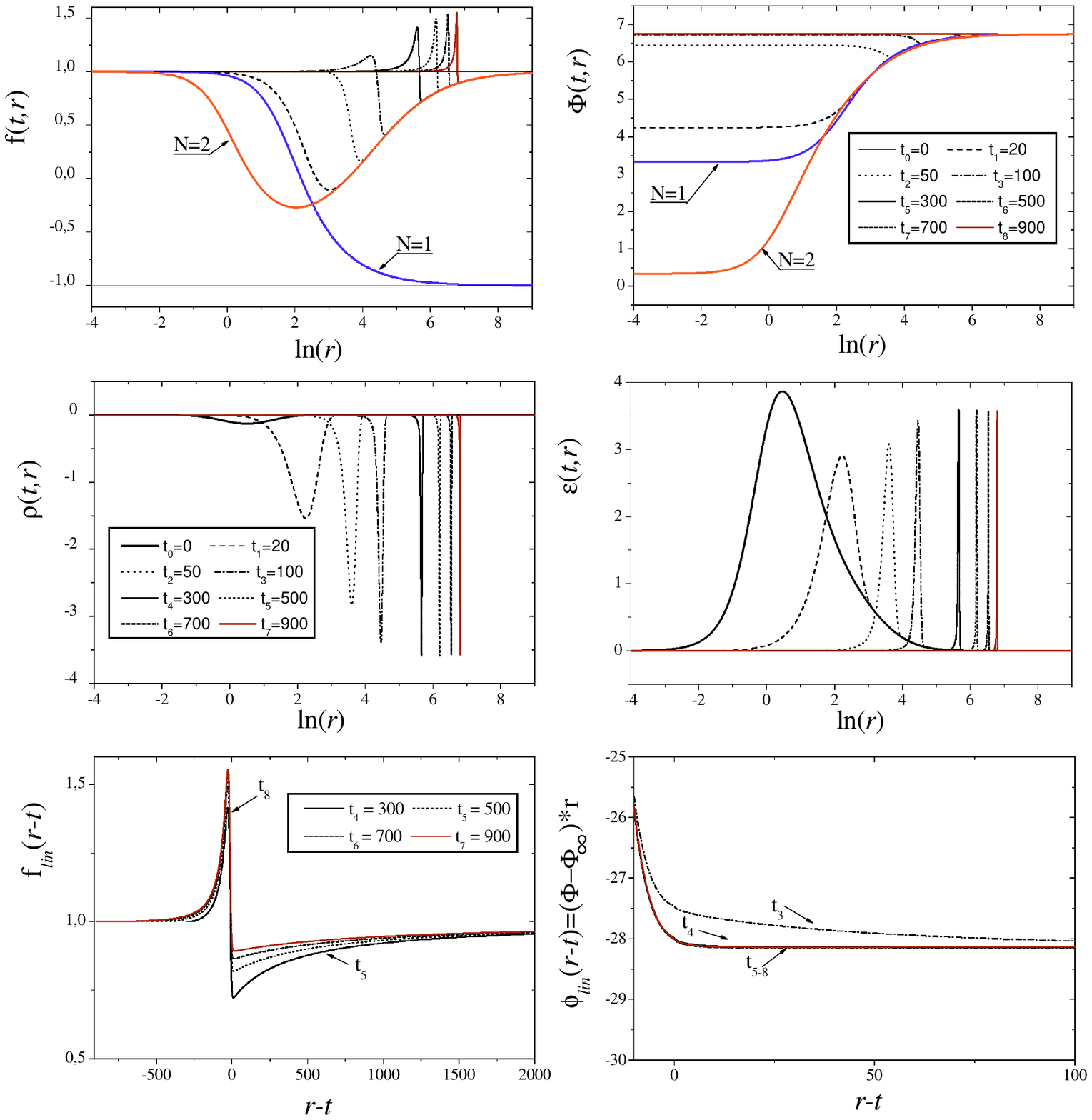} 
        \vspace*{-8.5cm} 
    \caption{
The decay of the $N=2$ static solution along its main unstable eigenmode with $\lambda_2^1=-7.5382 \times 10^{-2}$ and $\epsilon=+0.1$, which leads to scattering. The YM $f(t,r)$ (top, left) and the dilaton  $\Phi(t,r)$ (top, right) profiles are displayed at various moments of time versus the logarithm of the $r$. The radial momentum density function $\rho(t,r)$ (middle, left) and the energy density function $\epsilon(t,r)$ (middle, right) are shown at the same moments of time. The asymptotical profiles of the YM function $f_{lin}(t,r)$ (bottom, left) and the profiles of the normalized dilaton function $\phi_{lin}(t-r)=(\Phi(t,r)-\Phi_{\infty})\times r$ ($\Phi_{\infty}=6.7478$ is the asymptotical value of the dilaton function for the $N=2$ static solution)  (bottom, right) are also displayed versus the retarded null coordinate $r-t$.
}
\end{figure}


\newpage

\begin{figure}[!ht] 
\label{fig:5}
 \includegraphics[width=14cm]{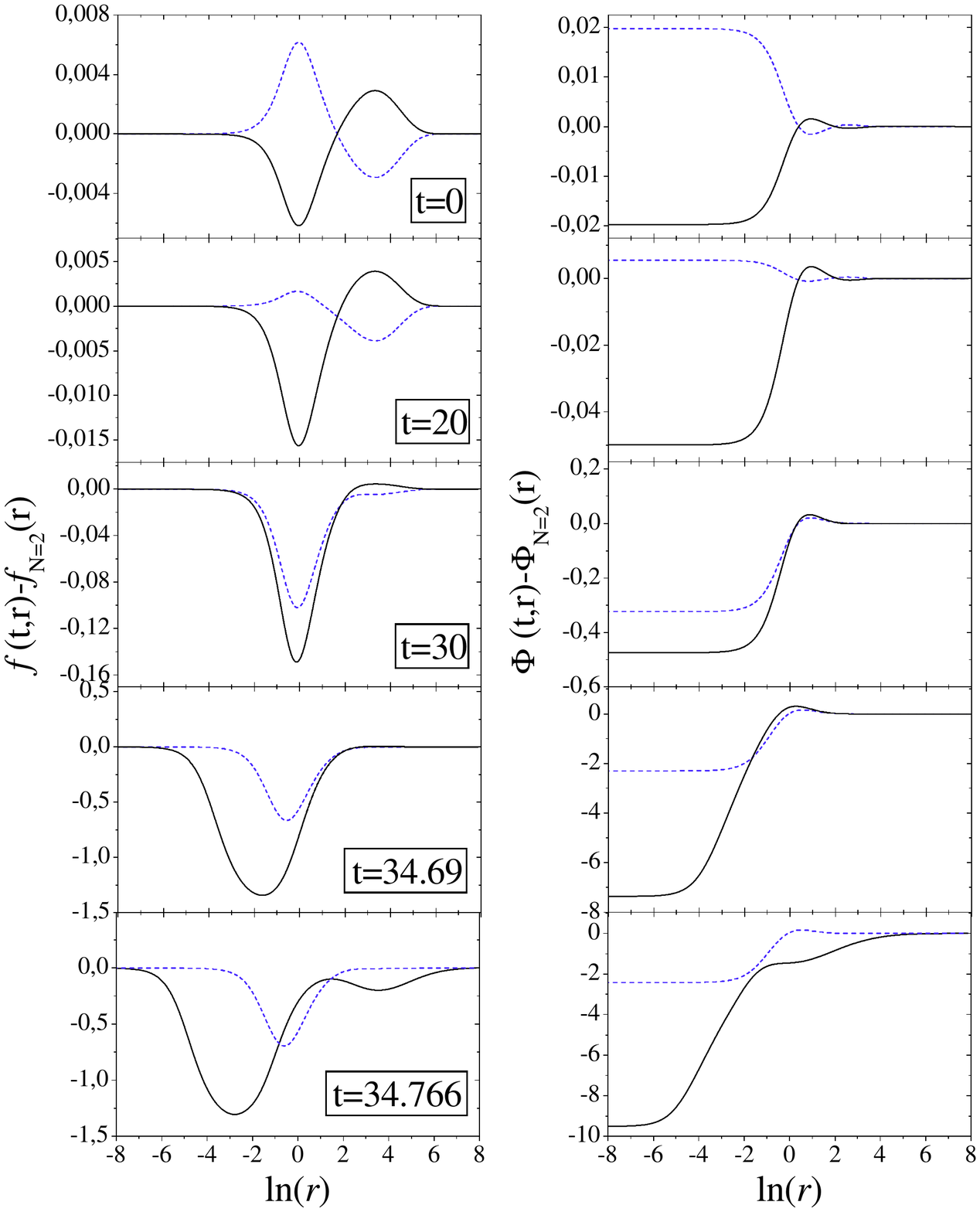} 
     \vspace*{-2cm} 
     \caption{
The decay of the $N=2$ static solution along its second unstable eigenmode with $\lambda_2^2=-7.5382 \times 10^{-2}$ and $\epsilon= \pm 0.5$.
 The deviations from the static $N=2$ solutions are shown for the YM $f(t,r)-f_{N=2}(r)$ (left) and for the dilaton $\Phi(t,r)-\Phi_{N=2}(r)$ (right) functions at various moments of time at the beginning of evolution. Solid lines correspond to the solution with $\epsilon=+ 0.5$, dashed lines - to the solution with $\epsilon=-0.5$.
}
\end{figure}

\newpage
\begin{figure}[!ht] 
    \includegraphics[width=14cm]{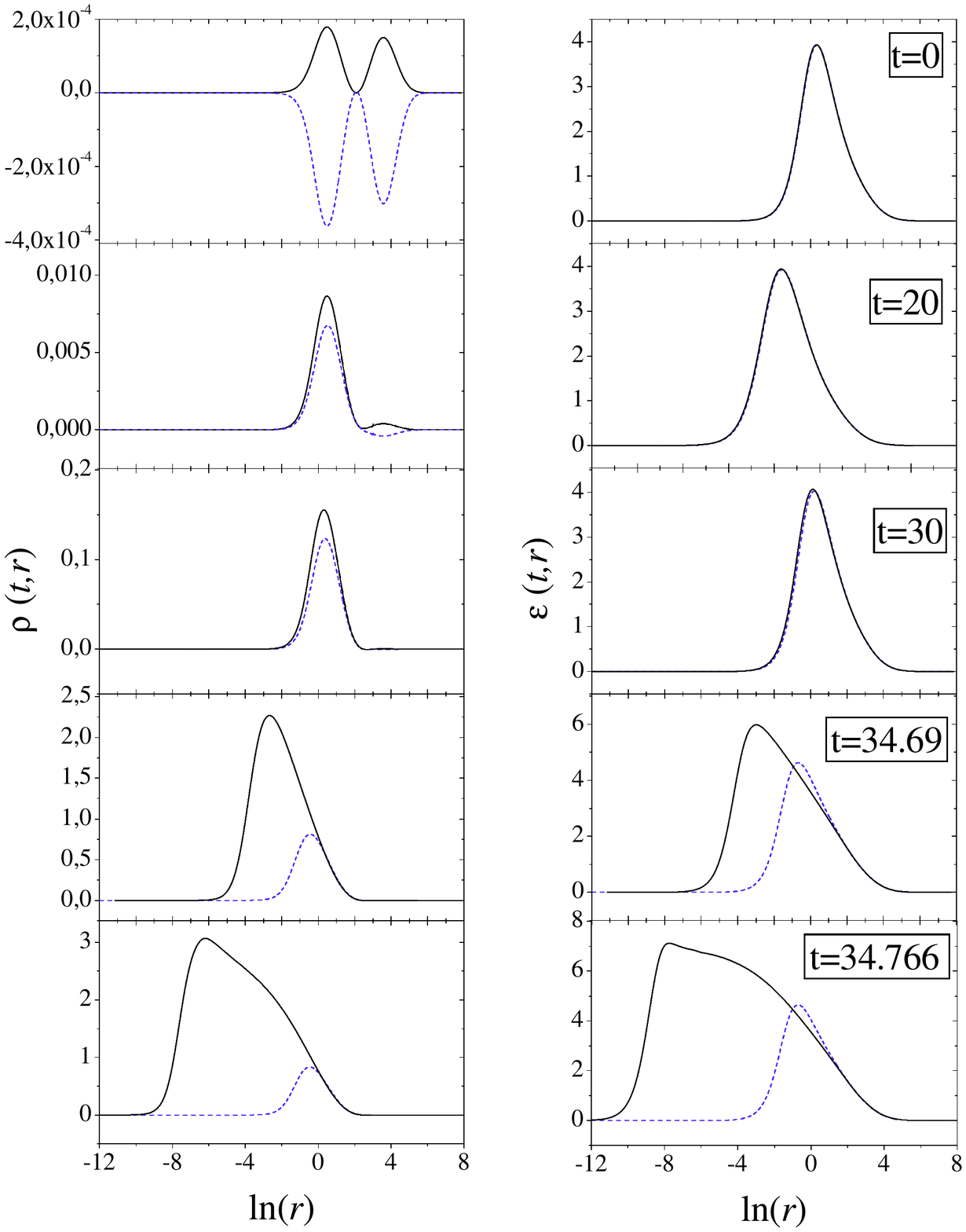} 
         \vspace*{-2cm} 
     \caption{The decay of the $N=2$ static solution along its second unstable eigenmode with $\epsilon=+0.5$ (solid lines) and $\epsilon=-0.5$ (dashed lines).  The radial momentum density function $\rho(t,r)$ (bottom, left) and the energy density function $\epsilon(t,r)$ (bottom, right) are displayed
at the same moments of time as in the previous Fig.8.
}
\end{figure}

\newpage


\newpage

\begin{figure}[!ht] 
 \label{fig:7}
    \includegraphics[width=17cm]{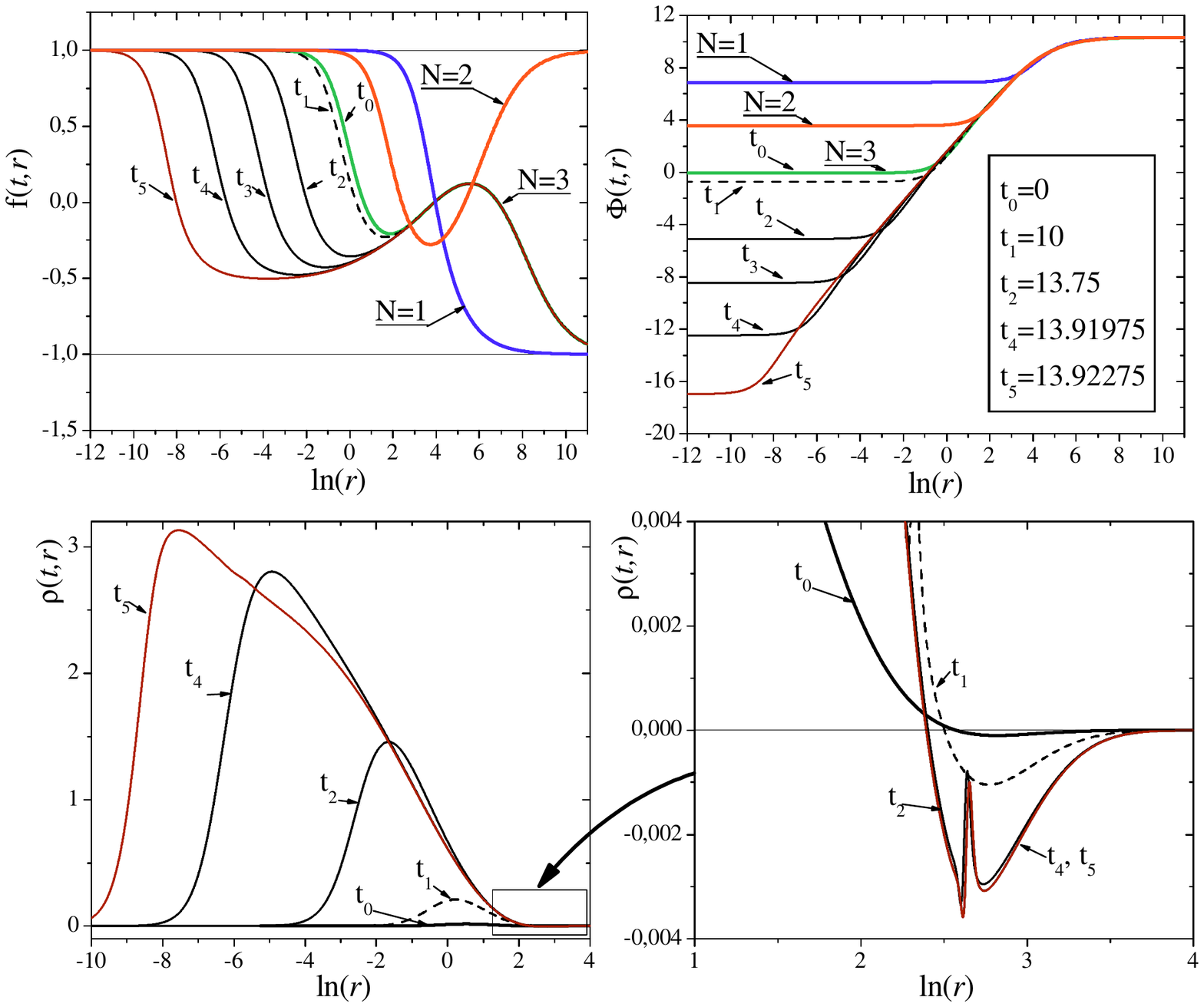} 
         \vspace*{-11cm} 
      \caption{
The decay of the $N=3$ static solution along its main unstable eigenmode 
with $\lambda_2^1=-4.9346 \times 10^{-2}$ and $\epsilon=-0.2$, which leads to blowup. The YM $f(t,r)$ profiles (top, left) and the dilaton profiles $\Phi(t,r)$ are plotted at various moments of time. The radial momentum density function $\rho(t,r)$ (bottom: left, right) is displayed at the same moments of time. The negative values of function $\rho(t,r)$ (bottom, right) indicate an outgoing energy flux, which can be detected at spatial infinity. 
}
\end{figure}

\end{document}